\documentclass{JHEP3}

\usepackage{amsfonts}
\usepackage{amsmath}
\usepackage{amssymb}
\usepackage{amsmath,amssymb,epsfig}
\numberwithin{equation}{section}
\usepackage{cite}

\newcommand{\be}{\begin{equation}}
\newcommand{\bea}{\begin{eqnarray}}
\newcommand{\eea}{\end{eqnarray}}
\newcommand{\ba}{\begin{array}}
\newcommand{\ea}{\end{array}}
\newcommand{\ee}{\end{equation}}

\newcommand{\bchi}{{\mbox{\boldmath $\chi$}}}
\newcommand{\pa}{\partial}

\newcommand{\cN}{\mathcal{N}}

\newcommand{\tzeta}{{\tilde\zeta}}
\def\bse{\begin{subequations}}
\def\ese{\end{subequations}}

\title{Supersymmetric black rings and non-linear sigma models }

\author{Yi-Xin
Chen and Yong-Qiang Wang \\
Zhejiang Institute of Modern Physics, Zhejiang University\\
 Hangzhou 310027, P. R. China\\
 E-mail: \email{yxchen@zimp.zju.edu.cn}, \email{wangyongqiangyueyuan@gmail.com}}

\abstract{ In this paper we investigate the non-linear sigma model
arising in the reduction of D = 5 supergravity to D = 3, and present
the application of this sigma model to supersymmetric black ring
solutions in five-dimensional minimal supergravity. With the ansatz
of stationary solutions with $R \times U(1)\times U(1)$ isometry, we
obtain a two-dimensional Lagrangian corresponding to geodesic motion
of a string-like object on the coset $G_{2(+2)}/SO(4)$, and study
the algebra of conserved charges and supersymmetry constraints of
supersymmetric black rings. We also obtain the semi-classical wave
function of supersymmetric black rings. }

\begin{document}

\section{Introduction}
The study of black rings in high dimensions has been a subject of
great interest over the past few years. As is well known, the
horizon topology of  black holes in high dimensional spacetime is
not unique.  The discovery of a new BH phase : a rotating black hole
solution with horizon topology $S^1\times S^2$ and carrying angular
momentum along the $S^1$, is called as black ring.
 It was initiated by Emparan and Reall who focused on the  solution of the five-dimensional vacuum
Einstein equations \cite{4},
then extended to include electric charge \cite{Elvang:2003yy}.
Remarkably, the discovery of a supersymmetric black ring solution of
five-dimensional minimal supergravity was introduced in \cite{5}.
Several important developments are studied more recently in
\cite{6,7}. For the  review of black ring solutions, see
Refs.\cite{8}.

It is known that there exists a variety of solution generation
methods which were developed to derive these ``black ring''
solutions. These methods allow to find solutions possessing a
certain number of isometries. After reducing the high-dimensional
solutions to three dimensions with the corresponding  Killing
vectors, one can obtain non-linear sigma models in three dimensions,
which  are harmonic maps from a 3-dimensional
 base space to a $N$-dimensional target space $\mathcal {M}$ . The
target space is isomorphic to the coset $G/H$,  with  $G$ being the
isometry group of $\mathcal {M}$  and $H \in G$ the local isotropy
group. In general, the non-linear sigma model can be considering as
the effective theory of high-dimensional gravity model. In the
context of five-dimensional minimal supergravity with two commuting
Killing vectors: one time-like and one space-like Killing vectors,
one of which is hypersurface-orthogonal, it was shown that the
dimensional reduction to three dimensions leads to three-dimensional
non-linear sigma models coupled to gravity, with the moduli spaces
$G_{2(+2)}/SO(4)$ \cite{Bodner:1989cg,Cremmer:1999du,6t:89}  in the
case of a Lorentzian three-space. Several important application of
this model has been discussed in
\cite{Bouchareb:2007ax,Gal'tsov:2008nz,Clement:2007qy,Gaiotto:2007ag,Bergshoeff:2008be,Clement:2008qx,Gal'tsov:2008sh}.

Recently, Berkooz and Piolin in \cite{Berkooz:2008rj} addressed this
non-linear sigma model and further study  the stationary solutions
of five-dimensional minimal supergravity
 with the symmetry group of moduli space.
Especially, by imposing the ansatz of stationary solutions with
$SU(2)\times U(1)$ isometry, one obtain a one-dimensional Lagrangian
 corresponding to geodesic motion on the coset $G_{2(+2)}/SO(4)$.
 Analyzing
the algebra of conserved charges and supersymmetry constraints, the
authors also obtain the corresponding semi-classical radial wave
functions of black holes. The above work  is in the context of BPS
black hole solutions, naturally,  the authors of
\cite{Berkooz:2008rj} also hope to try to extend these methods to
include the stationary solutions with $U(1) \times U(1)$ isometries,
which is relevant to black ring solutions.

It is the aim of this article to further study the application of
the non-linear sigma model to  black ring solutions in five
dimensional supergravity. Inspired by \cite{Berkooz:2008rj}, we
study the five-dimensional black rings in 5D supergravity with
$U(1)$ isometry by dimensional reduction to three dimensions. With
the ansatz of stationary solutions with $R \times U(1)\times U(1)$
isometry, we obtain a two-dimensional Lagrangian with the
coordinates $r$ and $\theta$, which describes  the geodesic motion
of the string-like on coset space $G_{2(+2)}/SO(4)$, and study the
algebra of conserved charges  and supersymmetry constraints of
supersymmetric black rings. Furthermore, by solving the
supersymmetry constraints, we also obtain the semi-classical wave
function of supersymmetric black rings, which describes that the
motion of a  string-like object can be represented as a
semi-classical wave function.

This paper is organized as follows. In the next section, we briefly
review the supersymmetric solutions of $\mathcal {N}=1$ supergravity
and specialized to the case of black rings in the Gibbons-Hawking
spaces. In section \ref{weqq} we give a brief review about the
non-linear sigma model arising in the reduction of D = 5
supergravity to D = 3, and present the application of this sigma
model to $R \times U(1)\times U(1)$ symmetric solutions in
five-dimensional minimal supergravity. In section \ref{charges} we
 compute the conserved
charges and supersymmetry constraints of supersymmetric black rings,
and obtain the semi-classical wave function of supersymmetric black
ring.  The last section is devoted to discussions.

\section{Supersymmetric black rings in  $D=5$ , $\mathcal {N}=1$ supergravity}\label{intro}
 In the present
section we briefly present a review of the five dimensional low
energy  supergravity theory. From the view of M-theory, the
compactifications of M-theory on a Calabi-Yau manifold $CY_3$ lead
to the five dimensional supergravity coupled to an arbitrary number
of vector multiplets. This model is usually studied in the context
of real or very special geometry. Relevant references  can be found
in \cite{11, 12, 13,14}. We will adopt the same conventions as
\cite{firstorder}.

At the two-derivative level, the bosonic part of $\mathcal {N}=1$
$D=5$ ungauged supergravity coupled to $n-1$ abelian vector
multiplets with scalars $\phi^i, i= 1,...,n-1$, is read as\bea
\label{action}
 S = {1 \over 16 \pi G_{5}} \int   R \star 1 -G_{IJ} \big( F^I
 \wedge \star F^J + dX^I \wedge
\star dX^J \big)
  -{C_{IJK} \over 6}  F^I \wedge F^J \wedge A^K,
\eea which the two-form field strength $F^I = dA^I$, the real
scalars $X^I= X^I(\phi^i)$ , $I,J,K$ = $1 , ... ,n$, are defined by
the cubic equation: $\frac{1} { 6} C_{IJK} X^I X^J X^K=1$ and the
constants $C_{IJK}$ are the topological intersection numbers and
symmetric on $IJK$. The metric $g_{ij}$ on the scalar manifold with
the coordinates $\phi_i$ is given by  \be g_{ij}=G_{AB}\partial_i
X^I
\partial_j X^J \, ,\ee
where the notation $\partial_i = \frac{\partial }{\partial
\phi^i}$\,. It is convenient to define  \be \label{notation}
 X_I \equiv {1 \over 6}C_{IJK} X^J X^K\, ,  \,\, \Rightarrow X_I X^I=1\,,\ee and so \be X^I\partial_i X_I =
\partial_i X^I X_I=0\, .\ee Thus one can express $G_{IJ}$ in terms of $X_I$ through the
equations: \be G_{IJ} = {9 \over 2} X_I X_J -{1 \over 2}C_{IJK} X^K
\, . \label{matrix} \ee

Using the above Eqs. (\ref{notation}) and (\ref{matrix}) , one can
follows that the relation between the derivative of the special
coordinate $X_I$  and that of the dual coordinate  $X_I$ is given by
\be X_A= \frac{2}{3}G_{AB} X^B \, ,\,\,\,\,\,X^A= \frac{3}{2}G^{AB}
X_B \, ,\ee  and
\be
\partial_i X_A = -\frac{2}{3}G_{AB}\partial_i X^B \, , \,\,\,
\partial_i X^A = -\frac{3}{2}G^{AB}\partial_i X_B \,. \label{modulisca}\ee

\subsection{Time-like case and BPS equations}
We are interested in solutions preserving some supersymmetry. It is
well known that in general  a spacetime symmetry can be represented
by a generating Killing vector field. To obtain the stationary
solutions, we need find the corresponding Killing vector.
All supersymmetric solutions of five-dimensional  supergravity
\cite{15,Gauntlett:2003fk} imply the existence of a non-spacelike
Killing vector field. By assuming that in a region the Killing
vector field $V=\partial/\partial t$ is time-like , the
 line element of five dimensions and  field strengths $F^I$ can be
 written as
\begin{equation}\label{5D}
 ds_5^2 = -f^2 (dt+\omega)^2 +f^{-1} ds_{{\cal M}_{4}}^2~, \,\,F^I = d\left[(fX^I(dt+\omega)\right] +\Theta^I~,
\end{equation}
where ${\cal M}_{4}$ is a four-dimensional hyper-K\"{a}hler
manifold. $f$ is a scalar function harmonic on ${\cal M}_{4}$,
$\omega$ and $\Theta^I$ are  a 1-form  and closed  self-dual 2-forms
, respectively.
In additional of the  supersymmetry transformation of the bosonic
fields ,
  the scalar
   $f$, one-forms $\omega$ and $\Theta^I$ on ${\cal M}_{4}$
are given by the following equations  named as ``BPS'' equations:
\be (\Theta^I)^- = 0~, \,\,\, \triangle_{{\cal M}_{4}} (f^{-1} X_I)=
{1 \over 6} C_{IJK} \Theta^J \cdot\Theta^K~,\,\,\, (d\omega)^+ =
-\frac{3}{2} f^{-1} X_I \Theta^I~, \label{bpsequa} \ee here, the
operator $\triangle_{{\cal M}_{4}}$ is  the Laplacian. We use the
superscripts ``$\pm $'' to denote  the self-dual and antiself-dual
part with respect to the base ${\cal M}_{4}$, and for 2-forms
$\alpha$ and $\beta$ on ${\cal M}_{4}$ we define $\alpha \cdot \beta
= \alpha^{mn } \beta_{mn}$, with indices raised by the matrix
$h^{mn}$ on ${\cal M}_{4}$.

\subsection{Black rings solutions in Gibbons-Hawking base spaces }
To  further analyze the solutions of the BPS equations
(\ref{bpsequa})  in more detail, we will concentrate on the
so-called Gibbons-Hawking base spaces. It has been shown that such
base manifold admits the existence of a Killing vector which can
preserves the hyper-K\"{a}hler structure. Assuming the existence of
the Killing vector $\partial_{ 5}$, the metric of the base space
${\cal M}_{4}$ in the Gibbons- Hawking coordinates is given by  \be
\label{basespace} ds_{{\cal M}_{4}}^2 = H^{-1} (dx^5 + \chi)^2 +H
\delta_{ij} dx^j dx^j \,, \ee where $\chi=\chi_i dx^i$ ($i, j = 1,\,
2, \, 3$) ,  $H$ and $\chi$ are independent of $x^5$ and can be
solved explicitly, $\chi$ is determined by $\nabla \times \bchi =
\nabla H$, which implies that $H$ is a harmonic function on the
Euclid space $\mathbb{E}^3$.
We introduce one-forms $\eta^I$ and  $\Theta^I = d\eta^I $. It is
convenient to set
\begin{subequations} \label{BPSflow5d1}
\begin{align}
\omega &= \omega_5 (dx^5 + \chi)+ {\hat{\omega}_4} \, ,\\
\eta^I &= \eta^I_5 (dx^5 + \chi)+ \hat{\eta}_4^I \, ,
\end{align}
\end{subequations}
where $\hat{\omega}_4=\omega_{4i} dx^i,~\hat{\eta}_4^I=\eta^I_{4i}
dx^i$. Solving the BPS equations , we obtain a general
supersymmetric solution  in terms of  $2n+2$ harmonic functions $H$
, $K^I$ , $L_I$ and $B$ on $\mathbb{E}^3$\cite{6}: \bea \nabla
\times {\hat{\omega}_4} &=& H \nabla B - B \nabla H
+\frac{3}{4}(L_I\nabla K^I- K^I\nabla L_I)
\,,\label{addicon}\\
\omega_5 &=& - {1 \over 48}H^{-2}C_{IPQ} K^I K^P K^Q-{3 \over 4}
H^{-1} L_I K^I +B \,,\label{formeta}\\f^{-1} X_I &=& {1 \over 24}
H^{-1} C_{IPQ} K^P K^Q +L_I. \eea

It is well known that $H$ determines the Gibbons-Hawking base, such
as three examples of Gibbons-Hawking metrics: flat space ($H = 1$ or
$H = 1/|{\bf x}|$), Taub-NUT space ($H = 1 + 2M/|{\bf x}|$) and the
Eguchi-Hanson space ($H = 2M/|{\bf x}| + 2M/|{\bf x-x_0}|$)
\cite{15}. In order to get the black ring solution with the horizon
topology of $S_2$$\times$$S_1$,
 it
is convenient to take the base space $\cal M$ to be flat space
$\mathbb{E}^4$ with metric  \be \label{hypkahl} ds^2_{\mathbb{E}^4}
= H^{-1} (d \psi+\chi)^2+ H(dr^2+r^2 \big[d \theta^2 + \sin^2 \theta
d \phi^2 \big])\, , \ee where $H=1/|{\bf x}|\equiv 1/r$ and $\chi =
\cos\theta d\phi$, which satisfies $\nabla\times {\bchi}=\nabla H$.
The range of the angular coordinates are $0<\theta<\pi$, $0<\phi<2
\pi$ and $0<\psi<4\pi$. Thus, the multi-black rings solutions is
given
by \cite{6} \bea \label{eqn:newgensol} K^I &=& \sum_{i=1}^M q^I{}_i h_i \,, \nonumber \\
L_I &=& \lambda_I  +{1 \over 24} \sum_{i=1}^M (Q_{Ii} - C_{IJK}
q^J{}_i q^K{}_i) h_i \,, \nonumber \\ B &=& \frac{3}{4} \sum_{i=1}^M
\lambda_I q^I{}_i - \frac{3}{4}
 \sum_{i=1}^M \lambda_I q^I{}_i |{\bf{x}}_i|h_i \,,
\eea where $h_i$ are harmonic functions in $\mathbb{E}^3$ centred at
${\bf x}_i$, $h_i=1/|{\bf x}-{\bf x}_i|$, and $Q_{Ii}$, $q^I_i$ and
$\lambda_I$ are constants.

Next,  let us consider the special case of minimal $\mathcal {N}$ =
1 supergravity in five dimensions, which was first constructed in
\cite{5} and will be main background in our subsequent analysis. In
this model, a single supersymmetric black ring sitting at   ${\bf
x}_0 = (0, 0, -R)$ , a distance $R$ along the negative $z-$axis of
the three-dimensional space,
has \bea \label{spm1} f^{-1} &=& H^{-1}K^2 + L= 1 + \frac{Q-q^2}{4\Sigma} + \frac{r \,q^2}{4\Sigma^2} , \nonumber \\
\omega_5 &=& H^2K^3 + \frac{3}{2} H^{-1}K L + M = - \frac{r^2
q^3}{8\Sigma^3 } - \frac{3 r \, q}{4
\Sigma}(1+\frac{Q-q^2}{4\Sigma})+\frac{3q}{4}- \frac{3q\,R }{4\Sigma} ,\nonumber\\
K&=&-\frac{q}{2\Sigma},\,\,\quad L=1 +
\frac{Q-q^2}{4\Sigma},\,\,\quad B=\frac{3q}{4}- \frac{3q\,R
}{4\Sigma}.\,
 \eea
In addition, the gauge potential are given by  \be
\label{spm2}\nabla \times {\hat{\eta}_4} = -\nabla (H \eta_5 )
,\,\,\,
 \eta_5 = -\frac{r \, q}{2 \Sigma}+\frac{q}{2}, \ee
here $\Sigma=|{\bf x}-{\bf x}_0|=\sqrt{r^2+R^2+2 R \,r \cos\theta}$.
Noticed that in this special case  we can drop the
(sub-)superscripts $``I"$ in (\ref{eqn:newgensol}).

\section{Non-linear sigma
model and $R \times U(1)\times U(1)$ symmetric solutions
}\label{weqq} In this section, first, we will give  a brief review
about the non-linear sigma model arising in the reduction of D = 5
supergravity to D = 3. This is the starting point of that we can
present the application of this sigma model to supersymmetric black
ring solutions in five-dimensional minimal supergravity.
In order to obtain a non-linear sigma model, we proceed in the
following two steps:

(i) Reducing the five-dimensional space-time to four-dimensional
Euclidean space. Assuming the existence of a time-like Killing
vector $\partial_{\,t}$, the five-dimensional metric and gauge
fields can be taken the forms of (\ref{5D}). One can compactify
along time direction and reduce to 4D  spaces.
 This leads to four-dimensional Euclidean (instead of
Minkowskian)  $\cN=2$ supergravity , coupled to $n+1$ vector
multiplets, which has been studied in \cite{Cortes:2003zd}.

(ii) Further reduction to three dimensions with a space-like Killing
vector $\partial_{\,5}$. With this ansatz, four-dimensional
hyper-K\"{a}hler base $ds_{{\cal M}_{4}}^2$ can be written as the
form of (\ref{basespace}), and the gauge fields are read as
(\ref{BPSflow5d1}). The equations of motion for the gauge fields
$\hat{\eta}_4^\Lambda$ and $\chi$ can allow to define the dual
scalars $\tilde{\eta}_\Lambda$ and $\sigma$. The dual scalars for
$\hat{\eta}_4^\Lambda$ and $\chi$ are defined by{\footnote{Note that
here the reduction is from four-dimensional Euclidean space to three
dimension. Considering analytic continuation, we need perform an
wick rotation $\eta_5^\Lambda$ $\rightarrow$ $i \eta_5^\Lambda$. So
the dual equations have some difference comparing with the equations
in  the appendix of \cite{Gaiotto:2007ag}.}}:
\begin{subequations} \label{dualtwo}
\begin{align}
\label{DefinitionDualScalarOmega} d\tilde{\eta}_\Lambda &= i  e^{2U}(Im\mathcal{N})_{\Lambda\Sigma}\boldsymbol{\star}(d\hat{\eta}_4^\Sigma+\eta^\Sigma_5 d\chi)+(Re\mathcal{N})_{\Lambda\Sigma}d\eta^\Sigma_5 , \\
\label{DeDualScalarOmega}d\sigma &= i e^{4U}\boldsymbol{\star} d
\chi+(\eta^\Lambda_5 d\tilde{\eta}_\Lambda-\tilde{\eta}_\Lambda
d\eta^\Lambda_5 ),
\end{align}
\end{subequations}
where $e^{2U}= H^{-1}$,  which is defined as harmonic on the Euclid
space $\mathbb{E}^3$ in (\ref{basespace}),
$\mathcal{N}_{\Lambda\Sigma}$ is complex symmetric matrix on
symplectic sections. Thus, one reduces the five-dimensional
Lagrangian to three dimensions and obtains $\cN=4$ supergravity
coupled to a non-linear sigma model \cite{deWit:1992up} \be
\mathcal{L}_3 = -\frac12 R -\frac12 G_{ab} \pa_i\varphi^a
\pa^i\varphi^b  , \ee here,$i=1,2,3$ and the target space with the
coordinates  $\varphi^a=\{U,z^I,\bar z^I,\eta_5^\Lambda,
\tilde{\eta}_\Lambda,\sigma\}$, which are the moduli fields , and
$G_{ab}$ is the moduli space metric.

So far we have reduce the $D=5$ $\mathcal {N}=1$ supergravity to
$D=3$ gravity coupled to a non-linear sigma model  with the $R\times
U(1)$  isometry of the stationary solutions.  In order to study the
black ring solutions,  one need  impose an extra $U(1)$ group of
isometries to  supersymmetric stationary solutions, and further
 reduce to two dimensions with  the other space-like Killing
vector $\partial_{\,\phi}$. Taking the base space ${\cal M}_{4}$ to
be flat space $\mathbb{E}^4$ , one can obtain three-dimensional
metric as \be ds_3^2= N^2(\rho, \theta)dr^2+r(\rho,\theta )^2 \big[d
\theta^2 + \sin^2 \theta d \phi^2 \big], \ee with this metric, the
stationary solutions have a isometry group $R\times U(1)\times
U(1)$. Upon reduction along the $\phi$ direction, the Lagrangian is
written as
\begin{eqnarray}
    \label{stationarySBPS} \mathcal{L}_2 &=& \frac{2 \sin\theta}{N^2 r^2} \biggl[  - N^3
r^{2}_{\theta} +r
N^3 \csc\theta(\cos\theta \,r_{\theta}+ \sin\theta r_{\theta\theta})  \nonumber\\
    && +r^2 N(-N^2+ N \cot\theta \, N_{\theta}+
    NN_{\theta\theta}+r^2_{\rho}) \nonumber\\
    && \left.
    + r^3\left(-2r_{\rho}N_{\rho}+2 N r_{\rho\rho}   \right) \right.
    \biggr]- \frac{ \sin\theta \, r^2}{N}\,g_{ab}\,\varphi_\rho^{a} \varphi_\rho^{b}+ N \sin\theta \,g_{ab}\,\varphi_\theta^{a} \varphi_\theta^{b}
    \;.
\end{eqnarray}
 Thus we obtain the two-dimensional
effective Lagrangian with the coordinates $\rho$ and $\theta$, which
describes  the geodesic motion of the string-like object on coset
space. Remarkably, we can observe that effective Lagrangian
(\ref{stationarySBPS})  is more complicated than the case of $R
\times U(1)\times SU(2)$ symmetric solutions in
\cite{Berkooz:2008rj}.

\subsection{Algebra structure of coset space}
In the following part of our paper,  we will study a more
interesting model: minimal $\mathcal {N}$ = 1 supergravity in five
dimensions. In this background, we reduce the theory down to three
dimensions following the above method, and obtain  the corresponding
moduli space, which is the coset $G_{2(+2)}/SL(2,R) \times SL(2,R)$.
Relevant references can be found
in\cite{Bodner:1989cg,Cremmer:1999du,6t:89}. Following the
conventions of \cite{Berkooz:2008rj,Gunaydin:2007qq}, the coset
space $G_{2(+2)}/SL(2,R) \times SL(2,R)$ has metric:
\be \label{dsm3} ds_{\mathcal{M}_{3D}}^2 =
\epsilon_{{\alpha}{\beta}} \, \epsilon_{AB}\,
 V^{{\alpha}A}\otimes
V^{{\beta}B} = u \,\bar u + v \, \bar v + e \, \bar e + E \, \bar E
\ , \ee with the quaternionic vielbein \cite{Ferrara:1989ik} \be
\label{quatviel} V^{\alpha A} =
\begin{pmatrix} \bar u & \bar  v \\ \bar  e & \bar E \\ E &  e \\  v & u
\end{pmatrix}\ ,\quad
\ee where $\epsilon_{{\alpha}{\beta}}$ and $\epsilon_{AB}$ are the
anti-symmetric tensors, $\alpha=1,2$ and $A= 1,2,3,4$ , with
one-form entries defined as \cite{Gunaydin:2007qq} \bse
\label{quatvielg2} \bea u &=& \frac{e^{-U}}{ 2\sqrt{2}\,
\tau_2^{3/2}} \left( d\tzeta_0 + \tau\, d\tzeta_1
+ 3\, \tau^2 \,d\zeta_1 - \tau^3 \,d\zeta_0 \right),\\
v &=& dU - \frac{i}{2} e^{-2U} ( d\sigma - \zeta_0 d\tzeta_0
-\zeta_1 d\tzeta_1
+  \tzeta_0 d\zeta_0 +\tzeta_1 d\zeta_1 ),\\
e &=& -\frac{\sqrt{3}}{2\tau_2} d\tau  ,\\
E &=& -\frac{e^{-U}}{2\sqrt{6}\,\tau_2^{3/2} } \left( 3 d\tzeta_0 +
d\tzeta_1\, (\tau + 2\bar\tau) +3 \bar\tau\, (2\tau+\bar\tau)
\,d\zeta_1 -3 \tau\bar\,\tau^2\, d\zeta_0 \right) ,\eea \ese where
 the bar denotes complex conjugate and $\tau\equiv \phi^1+i f \equiv
\tau_1+i\tau_2$.

After reduction to 3D non-linear sigma model with the coset
$G_{2(+2)}/SL(2,R) \times SL(2,R)$ in the context of minimal
$\mathcal {N}=1$ supergravity, one can further perform  reduction to
two dimensions along the $\phi$ direction. The two-dimensional
Lagrangian (\ref{stationarySBPS}) is written as
\begin{eqnarray}
    \label{effectiv} \mathcal{L}_2 &=& \frac{2 \sin\theta}{N^2 r^2} \biggl[  - N^3
r^{2}_{\theta} +r
N^3 \csc\theta(\cos\theta \,r_{\theta}+ \sin\theta r_{\theta\theta})  \nonumber\\
    && +r^2 N(-N^2+ N \cot\theta \, N_{\theta}+
    NN_{\theta\theta}+r^2_{\rho}) \nonumber\\
    && \left.
    + r^3\left(-2r_{\rho}N_{\rho}+2 N r_{\rho\rho}   \right) \right.
    \biggr]- \frac{ \sin\theta \, r^2}{N}\left( u
\,\bar u + v \, \bar v + e \, \bar e + E \, \bar E \right)_\rho
\nonumber\\ &&+ N \sin\theta \,\left( u \,\bar u + v \, \bar v + e
\, \bar e + E \, \bar E \right)_\theta \;,
\end{eqnarray}
where the terms $( u \,\bar u + v \, \bar v + e \, \bar e + E \,
\bar E )_\rho$ and $ \left( u \,\bar u + v \, \bar v + e \, \bar e +
E \, \bar E \right)_\theta $ correspond to the values along $\rho$
and $\theta$ direction respectively. Thus, we obtain a
two-dimensional Lagrangian with the coordinates $\rho$ and $\theta$,
which can describes the geodesic motion of the string-like object on
coset space $G_{2(+2)}/SO(4)$, comparing with one-dimensional
effective Lagrangian \cite{Berkooz:2008rj} describing the geodesic
motion of a particle.

\subsection{Killing vectors and conserved currents}
Now, we study the algebra of conserved charges  with the $R\times
U(1) \times U(1)$ isometry of the stationary solutions. It is well
known that in the case of the black hole solution, the conserved
charges correspond to Killing vectors, which have been computed in
\cite{Berkooz:2008rj}. By replacing the operator
$\partial_{\varphi^a}$ by the momentum $p_{\varphi^a}$ conjugate to
$\varphi^a$, we can obtain  the conserved charges associated with
the corresponding Killing vector. Such methods were first proposed
in \cite{Gunaydin:2005mx}, and later extended to more models in
\cite{Pioline:2006ni,Alexandrov:2008gh,Berkooz:2008rj,Gunaydin:2007qq}.
However, for the stationary solutions with the $R\times U(1) \times
U(1)$ isometry, following the same method, the results associated
with the corresponding Killing vectors are dependent on the angular
coordinate $\theta$. We know that the conserved currents correspond
to the Killing vectors, which usually depend on the some angular
coordinates. In order to obtain the total conserved charges of the
model, we need integrate the currents over the angular coordinates.

Next, following the method in \cite{Berkooz:2008rj}, we give more
detail on how to computer the conserved charge of the stationary
solutions with the $R\times U(1) \times U(1)$ isometry. Considering
the Lagrangian (\ref{stationarySBPS}), we can obtain  the conjugate
momentum $p_{\varphi^a}$ of $\varphi^a$: \bse  \bea \label{momentum}
 p_U &=& 4 \frac{r^2 \sin\theta}{N} U' ,\quad
 p_{\tau_1} = 3 \frac{r^2\sin\theta}{N \tau_2^2} \tau_1'\ ,\quad
p_{\tau_2} = 3 \frac{r^2 \sin\theta}{N \tau_2^2} \tau_2'\
,\nonumber\\
\label{wqwqw}p_\sigma  &=& \frac{r^2 e^{-4U} \sin\theta}{N}
(d\sigma+\tzeta^I d\zeta_I - \zeta^I d\tzeta_I )\ ,\eea \bea
 p_{\tzeta_0} &=& \frac{r^2  e^{-2U}\sin\theta}{2N \tau_2^{3}}\big(4 \,d\tzeta_0 + 2(\tau+\bar\tau)d\tzeta_1 +3(\tau+\bar\tau)^2 d\zeta^1
 -\frac{1}{2} (\tau+\bar\tau)^3 d\zeta^0       \big)  -\zeta^0 p_\sigma , \nonumber\\
 p_{\tzeta_1} &=& \frac{r^2  e^{-2U}\sin\theta}{2N \tau_2^{3}}\big(2(\tau+\bar\tau)d\tzeta_0 + \frac{2}{3}(\tau^2+\bar\tau^2+4\tau\bar\tau)d\tzeta_1
 \nonumber\\
 &&+\frac{1}{2}(\tau^3+\bar\tau^3+11\tau\bar\tau^2+11\bar\tau\tau^2)^2
 d\zeta^1 -\frac{1}{2} (\tau+\bar\tau)^3 d\zeta^0       \big)  -\zeta^0 p_\sigma , \nonumber\\
 p_{\zeta^0} &=& \frac{r^2  e^{-2U}\sin\theta}{2N
\tau_2^{3}}\big(-\frac{1}{2}(\tau+\bar\tau)^3 d\tzeta_0
-\tau\bar\tau(\tau+\bar\tau)^2 d\tzeta_1
 \nonumber\\
 &&+4\tau^3\bar\tau^3
 d\zeta^0 -6 (\tau\bar\tau)^2(\tau+\bar\tau) d\zeta^1       \big)  +\tzeta_0 p_\sigma , \nonumber\\
p_{\zeta^1 } &=& \frac{r^2  e^{-2U}\sin\theta}{2N
\tau_2^{3}}\big(3(\tau+\bar\tau)^2 d\tzeta_0 +
\frac{1}{2}(\tau^3+\bar\tau^3+11\tau\bar\tau^2+11\tau^2\bar\tau)d\tzeta_1
 \nonumber\\
 &&+6(\tau^3\bar\tau+\bar\tau^3\tau+4\tau^2\bar\tau^2)
 d\zeta^1 -6 (\tau^2\bar\tau^3+\tau^2\bar\tau^3) d\zeta^0       \big)  +\tzeta_1 p_\sigma
 .\eea \ese

  It is well known that the Killing vectors
associated to the right-action of $G_{2(+2)}$ can generate the Lie
algebra of $G_2$, and  the 14-dimensional Lie algebra of $G_2$
consists of the  commutation relations. For details, see
\cite{Gunaydin:2007qq}. We replace the operator
$\partial_{\varphi^a}$  by the conjugate momentum $p_{\varphi^a}$
and rewrite the fourteen generators of $G_2$ as
 \bse  \label{monbr} \bea \label{noetherc1} E_k&=&p_{\sigma}, \quad E_{p^0}=
p_{\tzeta_0} - \zeta^0 p_{\sigma}  ,\quad
E_{q_0}= -p_{\zeta^0} - \tzeta_0 p_{\sigma} ,\nonumber\\
E_{p^1}&=& \sqrt{3}(p_{\tzeta_1} -  \zeta^1 p_{\sigma}) ,\quad
E_{q_1}= \frac{1}{\sqrt{3}}(-p_{\zeta^1} -  \tzeta_1 p_{\sigma}),\nonumber\\
H &=& -p_U -2 \sigma p_{\sigma} - \zeta^0 p_{\zeta^0} - \zeta^1
p_{\zeta^1} -   \tzeta_0 p_{\tzeta_0}- \tzeta_1 p_{\tzeta_1} ,\eea
\bea \label{noetherc2} Y_+&=& \frac{1}{\sqrt{2}}(p_{\tau_1} +
\zeta^0 p_{\zeta^1}
- 6\zeta^1 p_{\tzeta_1} - \tzeta_1 p_{\tzeta_0}),\nonumber\\
Y_0&=&-\frac12(2\tau_1 p_{\tau_1} + 2\tau_2 p_{\tau_2} - 3 \zeta^0
p_{\zeta^0}+ 3 \tzeta_0 p_{\tzeta_0}
- \zeta^1 p_{\zeta^1}+ \tzeta_1 p_{\tzeta_1}),\nonumber\\
Y_- &=& \frac{1}{3 \sqrt{2}} \left( 6 p_{\tau_2} {\tau_1} {\tau_2}+3
p_{\tau_1} \left(\tau_1^2-\tau_2^2\right)+9 p_{\tzeta_1}
{\tzeta_0}-9
   p_{\zeta^0} {\zeta^1}+2 p_{\zeta^1} {\tzeta_1}\right).
\eea \ese The other negative roots are too bulk to display.
It is obvious to see that the above Killing vectors depend on not
only the radial coordinate $r$ but also the angular coordinate
$\theta$. Noticed that
 in the case of the solution of black holes \cite{Berkooz:2008rj}, the Killing vectors
 are independent of $\theta$. Now, in the background of the stationary
solutions with the $R\times U(1) \times U(1)$ isometry, actually,
the Killing vectors we obtain in Eqs.(\ref{monbr})
 correspond to the conserved currents. These
currents describe the density of on the two-dimensional world sheet
of a string-like object.
  In order to obtain the conserved charges independent
of the coordinates, we should integrate these currents over the
angular coordinate $\theta$, i.e. a circle $S^1$ as the contour.

\section{Supersymmetric black rings} \label{charges}
In this section, we have an application of the above results to
supersymmetric black ring in minimal  D=5 , $\mathcal {N}=1$
supergravity.

\subsection{Conserved charges}
In  the last part of the section two , we have introduce the
solution of supersymmetric black ring. By making use of the result
in (\ref{spm1}) and (\ref{spm2}),
 we can write the coordinates  $\varphi^a$ of
the target space of $\sigma-$model as : \bse \be \label{bmpv}
\cN^2=H=\rho^{-1} ,\quad a^2=b^2=\rho\ ,\quad \tau_2 = -i \tau_1 = f
= \left(1 + \frac{Q-q^2}{4|{\bf x}-{\bf x}_0|} + \frac{\rho
\,q^2}{4|{\bf x}-{\bf x}_0|^2}\right)^{-1} \ ,\quad \ee \be \zeta^0
= \frac{1}{\sqrt{2}i}\left(- \frac{\rho^2 q^3}{8|{\bf x}-{\bf
x}_0|^3 } - \frac{3 \rho q}{4 |{\bf x}-{\bf
x}_0|}(1+\frac{Q-q^2}{4|{\bf x}-{\bf x}_0|})+\frac{3q}{4}- \frac{3qR
}{4|{\bf x}-{\bf x}_0|}\right)\ ,\quad \ee \be \zeta^1=
\frac{1}{\sqrt{2}}\left(-\frac{\rho \, q}{2 |{\bf x}-{\bf
x}_0|}+\frac{q}{2}\right), \ee  \ese where $Q$ is the electric
charge, equal to the ADM mass by the BPS condition, $q$ is the
dipole charge. In order to obtain the remaining coordinates of the
non-linear sigma model, one should solve the dual equations
(\ref{dualtwo}). In case of the black ring solutions, we can find
the $\tzeta_0$ and $\tzeta_1$ always take constant values by solving
Eq. (\ref{DefinitionDualScalarOmega}) . Moreover, analyzing the
connection between Eqs. (\ref{DeDualScalarOmega}) and (\ref{wqwqw}),
we can get $E_k = -i \sin\theta$.

Making use of the above results to solve
Eq.(\ref{DeDualScalarOmega}), we can obtain the remaining coordinate
of the non-linear sigma model, \bea \nonumber\sigma &=& -i \rho
-\frac{1}{\sqrt{2}i}\tzeta_0 (- \frac{\rho^2 q^3}{8|{\bf x}-{\bf
x}_0|^3 } - \frac{3 \rho q}{4 |{\bf x}-{\bf
x}_0|}(1+\frac{Q-q^2}{4|{\bf x}-{\bf x}_0|})-
\frac{3qR }{4|{\bf x}-{\bf x}_0|})\\
&&-\frac{1}{\sqrt{2}}\tzeta_1(-\frac{\rho \, q}{2 |{\bf x}-{\bf
x}_0|}) , \eea It is obvious to see that the eight coordinators
 of the target space  depend on not only the radial coordinate $r$ but also the angular
coordinate $\theta$, which is more complex than the case of black
hole. Remarkably, we  computer the first column of the viel-bein $V$
and obtain : \be \label{ueEv} \bar u = \bar e = E = v = 0 .\ee Thus,
we can observed that Eq.(\ref{ueEv}) takes the same result as that
of black hole in \cite{Berkooz:2008rj}. It is also shown that the
solutions (\ref{spm1}) and (\ref{spm2}) preserve the some
supersymmetry, which is consistent with the known facts.

Next, we should integrate the conserved currents in (\ref{monbr})
over the angular coordinate $\theta$, $0<\theta<\pi$,  i.e. a circle
$S^1$ as the contour . After a tedious calculation, we obtain the
conserved charges  for supergravity black ring: \bse
\label{noetherc} \bea
\int_{S^1}E_k &=&\int_{S^1} -i \sin\theta= - 2i ,\nonumber\\
\int_{S^1}E_{p^0}&=& \int_{S^1} p_{\tzeta_0} - \zeta^0 p_{\sigma}= 0
 ,\quad
\int_{S^1}E_{q_0}= \int_{S^1} -p_{\zeta^0} - \tzeta_0 p_{\sigma} = 4i \tzeta_0 ,\nonumber\\
\int_{S^1}E_{p^1}&=& \int_{S^1} \sqrt{3}(p_{\tzeta_1} -  \zeta^1
p_{\sigma})= 0  ,
\int_{S^1}E_{q_1}= \int_{S^1} \frac{1}{\sqrt{3}}(-p_{\zeta^1} -  \tzeta_1 p_{\sigma})= \frac{4i\tzeta_1}{\sqrt{3}} ,\nonumber\\
\int_{S^1}H &=&\int_{S^1} -p_U -2 \sigma p_{\sigma} - \zeta^0
p_{\zeta^0} - \zeta^1 p_{\zeta^1} -   \tzeta_0 p_{\tzeta_0}-
\tzeta_1 p_{\tzeta_1}\\\nonumber &=&- \frac{3}{\sqrt{2}}  q\tzeta_0
+ \sqrt{2} i q\tzeta_1, \eea \bea \int_{S^1}Y_+&=&
\int_{S^1}\frac{1}{\sqrt{2}}(p_{\tau_1} + \zeta^0 p_{\zeta^1}
- 6\zeta^1 p_{\tzeta_1} - \tzeta_1 p_{\tzeta_0})= \frac{3 i}{2 \sqrt{2}}Q ,\nonumber\\
\int_{S^1}Y_0 &=&\int_{S^1}-\frac12(2\tau_1 p_{\tau_1} + 2\tau_2
p_{\tau_2} - 3 \zeta^0 p_{\zeta^0}+ 3 \tzeta_0 p_{\tzeta_0}
- \zeta^1 p_{\zeta^1}+ \tzeta_1 p_{\tzeta_1})=0 ,\nonumber\\
\int_{S^1} Y_- &=& \int_{S^1} \frac{1}{3 \sqrt{2}} \left( 6
p_{\tau_2} {\tau_1} {\tau_2}+3 p_{\tau_1}
\left(\tau_1^2-\tau_2^2\right)+9 p_{\tzeta_1} {\tzeta_0}-9
   p_{\zeta^0} {\zeta^1}+2 p_{\zeta^1} {\tzeta_1}\right)\\\nonumber
   &=&
   -\frac{2\sqrt{2}}{3} i \tzeta_1^2, \\
 \int_{S^1}F_{p^0} &=&- \frac{3}{\sqrt{2}} q\tzeta_0^2  + \sqrt{2}
i q\tzeta_1\tzeta_0+ \frac{8 i \tzeta_1^3}{27}\ ,
\int_{S^1}F_{p^1}=\frac{3 i Q \tzeta_0  +
\sqrt{2}(-\frac{3}{2}q\tzeta_0+ i q \tzeta_1) \tzeta_1}{\sqrt3}\,,\nonumber\\
\int_{S^1} F_{q_1}&=&-\frac{2 i Q \tzeta_1}{\sqrt3}\ ,
\int_{S^1}F_{q_0}=-\frac{i}{2\sqrt2} q (12R + 3 Q- q^2)\ , \nonumber\\
\int_{S^1}F_k &=& -\frac{i}{24}\left[ 6 \sqrt2 q (12R + 3 Q- q^2)
\tzeta_0 + 3 (-\frac{3}{2}q\tzeta_0+ i q \tzeta_1)^2 + 4 Q
\tzeta_1^2 \right]\ .  \eea  \ese It is obvious to show that the
above results confirms the identification of $\int_{S^1} Y_+$ and
$\int_{S^1} F_{q_0}$ as the electric charge and  $J_\psi$,  one of
two angular momenta\cite{5}. Moreover, from the conserved charges
$\int_{S^1}E_{p^0}$ and $\int_{S^1}E_{p^1}$ taking the value of zero
,  it is well known that the black rings have only purely electric
charges.

In general, the angular momenta of  a supersymmetric black ring are
always unequal.  Two independent angular momenta $J_\phi$ and
$J_\psi$ correspond to the Killing vector $\partial_\phi$ and
$\partial_\psi$. Observing the results in (\ref{noetherc}), we only
find a conserved charge $\int_{S^1} F_{q_0}$ associating with the
angular momenta $J_\psi$. In order to identify the angular momenta
$J_\phi$, we need know the  dimensions of the fourteen Killing
vectors and find those conserved currents which have the same
dimensions with $F_{q_0}$. Choosing the dipole charge $q$ as the
base dimensions , thus, we obtain \be
[F_{q_1}]=[F_{p_0}]=[F_{p_1}]=[F_{q_0}]=[q]^3 .\ee Meanwhile,
considering that a supersymmetric black ring has two independent
angular momenta,  and  the term commutating with $F_{q_I}$ in
\cite{Gunaydin:2007qq} is only the generator $F_{q_J}$: \be
[F_{q_I},F_{q_J}]=0, \ee we can consider that the conserved charge
$\int_{S^1} F_{q_1}$ is the angular momenta $J_\phi$. Thus, using
the algebra of conserved charges, we computer the conserved charges
of supersymmetric black rings in minimal $\mathcal {N}=1$
supergravity and obtain the result consistent with \cite{5}.

\subsection{Hamilton-Jacobi equation
and the semi-classical  wave function of supersymmetric black rings}
In this subsection we give a discussion about the semi-classical
wave function of supersymmetric black rings. It is well known that
  the motion of a particle can be represented as  a wave function in  the formulation of  Hamilton-Jacobi
  equation
(HJE). We can introduce the function $S(x_1,...,x_N, t)$ called
 phase function and require that $S$ satisfy
HJE:
\begin{equation}\label{hje}
    H(x_1,...,x_N,\frac{\partial S}{\partial x_1},...,\frac{\partial
S}{\partial
    x_N},t)+\frac{\partial S}{\partial t}=0.
\end{equation}
 The conjugate momenta correspond to
the first derivatives of $S$ with respect to the generalized
coordinates: \be \label{cma} p_k = \partial_{x_k}S. \ee Thus  the
corresponding wave function can be written as:
\begin{equation}\label{hjewave}
    \Psi \propto \exp (i S).
\end{equation}

Now, using the same method in \cite{Berkooz:2008rj}, we can study
the supersymmetric constraints
 and further obtain the semi-classical wave function of supersymmetric black
 rings. In the above,
we know that the five constraints keep the supersymmetric properties
of the solution: \be \label{eee} \bar u = \bar e = E = v =
p_r+4\sin\theta = 0. \ee Considering the conjugate momenta
(\ref{cma}) and conserved currents \be Y_+ = \tilde Q(\rho,\theta)=
\frac{3}{\sqrt{2}}( i\rho^2\frac{d \tau_2}{\tau_2^2}+12\rho\zeta^1
d\zeta^1-6i(\zeta^1)^2) ,\quad E_k = \tilde K(\rho,\theta)=- i
\sin\theta ,\ee  the equations (\ref{eee}) can be solved. Meanwhile,
we obtain the phase function\be
\begin{split} \label{pf}
S_{\tilde Q,\tilde K,f} =& -4 \rho \sin\theta + i \tilde K e^{2U} +
\sqrt2 \tilde Q \bar\tau + \tilde K \left[ \sigma + \zeta^0 \tzeta_0
+ \zeta^1 \tzeta_1 + 6 \bar\tau \zeta_1^2 - 6 \bar\tau^2
\zeta^0\zeta^1
+2 \bar\tau^3 (\zeta^0)^2 \right]\\
&+ f \left( \tzeta_0 + \bar\tau \tzeta_1 +3 \bar\tau^2 \zeta^1 -
\bar\tau^3\zeta^0, \tzeta_1+ 6 \bar\tau \zeta^1 - 3 \bar\tau^2
\zeta^0 \right),
\end{split}
\ee with  $f$ is an arbitrary function of two variables.

Substituting  the phase function (\ref{pf}) into the wave function
(\ref{hjewave}), we find that the  solutions of  the semi-classical
approximation can be expressed as \be \Psi_{\tilde Q,\tilde K
,p^0,p^1} \propto \exp\left[ i S_{\tilde Q,\tilde K,0} + i p^0
\left( \tzeta_0 + \bar\tau \tzeta_1 +3 \bar\tau^2 \zeta^1 -
\bar\tau^3\zeta^0\right) + i p^1 \left( \tzeta_1+ 6 \bar\tau \zeta^1
- 3 \bar\tau^2 \zeta^0\right) \right] .\ee  Thus,
 we  obtain the semi-classical wave function of supersymmetric black
 rings, which is dependent of the coordinates $\rho$ and $\theta$. It is obvious to see that
  the motion of a  string-like object can be represented as  a  semi-classical wave
  function.
 It is noticed that
the wave function of supersymmetric black
 ring take the form analogous to that of black hole in five
 dimensions \cite{Berkooz:2008rj}. However, since the moduli fields $\varphi^a=\{U,z^I,\bar z^I,\eta_5^\Lambda,
\tilde{\eta}_\Lambda,\sigma\}$ are dependent on angular coordinate,
the wave function is more complex.

\section{Conclusion}
In this paper, with the $R\times U(1)$  isometry of the stationary
solutions, we reduce 5D supersymmetric black ring solutions to three
dimensions and obtain a non-linear sigma model. Further, adding the
extra $U(1)$ symmetry to the sigma model, we  obtain the
two-dimensional effective Lagrangian with the coordinates $r$ and
$\theta$, which describes  the geodesic motion of the string-like
object on the target space. Because the target space of the sigma
model is the symmetric space $G_{2(+2)}/SO(4)$, we can analyse the
Lie algebra of $G_{2(+2)}$ and study the algebra of conserved
charges and supersymmetry constraints of supersymmetric black rings.
In the context of black rings, the conserved charge is dependent of
coordinate $\theta$. In order to obtain the charges independent of
coordinates, we integrate these charge
 over $S^1$. After a tedious calculation, we obtain that
$\int_{S^1} Y_+$, $\int_{S^1} F_{q_1}$ and $\int_{S^1} F_{q_0}$
correspond  with  the electric charge $Q$ and  two angular momentums
$J_\phi$ and $J_\psi$ respectively. Moreover, analyzing the
supersymmetric constraints (\ref{eee}), we  obtain the
semi-classical wave function of supersymmetric black
 rings, which is dependent of the coordinate $r$ and $\theta$  and  describe that
 the motion of a  string-like can be represented as  a  semi-classical wave
  function.

\section*{Acknowledgement}

We would like to thank  Q. J. Cao, Y. J. Du, K. N. Shao and Y. Xiao
for useful discussions. The work is supported in part by the NNSF of
China Grant No. 90503009, No. 10775116, and 973 Program Grant No.
2005CB724508.


\end{document}